\documentclass[12pt]{aipproc}
\layoutstyle{6x9}

\def\kpipi{$D^+\rightarrow K^-\pi^+\pi^+$\ }
\def\kpp{$K^-\pi^+\pi^+$\ }
\def\ka14{$K^*_0(1430)$\ }
\def\d3pi{$D^+\rightarrow \pi^-\pi^+\pi^+$\ }
\def\ds3pi{$D_s^+\rightarrow \pi^-\pi^+\pi^+$\ }
\begin{document}

\title{Light Meson Physics from Charm Decays at Fermilab E791}

\author{Carla G\"obel~\footnote{For the E791 Collaboration.}}
{address={Instituto de F\'\i sica, Facultad de Ingenier\'\i a, 
Universidad de la Rep\'ublica, C.C.~30, CP~11300, Montevideo, Uruguay}}

\begin{abstract}
We present recent results on light mesons based on Dalitz plot 
analyses of charm decays from Fermilab experiment E791. Scalar mesons are found
to have large contributions to the decays studied, $D^+\to K^-\pi^+\pi^+$ and $D^+, D_s^+\to\pi^-\pi^+\pi^+$.
From the $K\pi\pi$ final state, we find good evidence for the existence of the light and broad 
$\kappa$ meson and we measure its mass and width. We also discuss recently published results 
on the 3$\pi$ final states, especially the measurement of the $f_0$ parameters and the evidence 
for the $\sigma$ meson from $D^+\to\sigma\pi^+$.  These results demonstrate the importance 
of charm decays as a new environment for the study of light meson physics.
\end{abstract}

\maketitle

\section{Introduction}

The decays of charm mesons are currently a new source of information for the study
of light meson spectroscopy, with the advantages of having well defined initial state
(the $D$ meson, a $0^-$ state with defined mass). This new information is complementary to that 
from scattering experiments and can be particularly relevant to the understanding
of the scalar sector.

Here we present preliminary results for the Dalitz-plot analysis of the Cabibbo-favored decay
\kpipi using data from Fermilab E791 experiment. We also present an overview of our results
for \ds3pi \cite{ds3pi} and \d3pi \cite{d3pi} Dalitz-plot analyses.
The E791 data was collected in 1991/92 from 500 GeV/c $\pi^-$-nucleon interactions. For details
see \cite{ref791}. 

For the \kpipi analysis, when we include all known $K\pi$ resonant channels plus a non-resonant 
(NR) contribution, we find that the NR decay is dominant. This is unusual in $D$ decays. Moreover,
the fit model has important discrepancies with respect to the data. By including an extra
scalar resonant state, with unconstrained mass and width, we obtain a fit which is substantially
superior to that without this state. The values for its mass and width are found to be
$797\pm 19\pm 42$ MeV/c$^2$ and $410\pm 43\pm 85$ MeV/c$^2$ respectively.
We refer to this state as the $\kappa$. The existence of such a state has been greatly discussed 
in the literature in recent years \cite{beveren}--\cite{cherry}.
We also obtain new measurements for the mass and the width of the \ka14 resonance:
$1459\pm 7\pm 6$ MeV/c$^2$ and $175\pm 12\pm 12$ MeV/c$^2$ respectively.

From our analysis of \ds3pi decays, we find that the dominant decay fraction comes
from $f_0(980)\pi^+$. We obtain new measurements for the $f_0(980)$ and $f_0(1370)$ masses
and widths. From \d3pi decays, we find that a model with only known $\pi\pi$ resonances plus
a NR channel is not able to describe the data adequately. We find strong evidence for the
presence of a light and broad scalar resonance, the $\sigma(500)$, the $\sigma\pi^+$ channel
being responsible for half of the decay rate. We measure the mass and the width of this scalar
meson to be $ 478^{+24}_{-23} \pm 17  $ MeV/$c^2$ and $ 324^{+42}_{-40}\pm 21$ MeV/$c^2$, 
respectively. 

\section{The \kpipi Dalitz-plot Analysis}

\begin{figure}[t]
\resizebox{8cm}{!}{\includegraphics{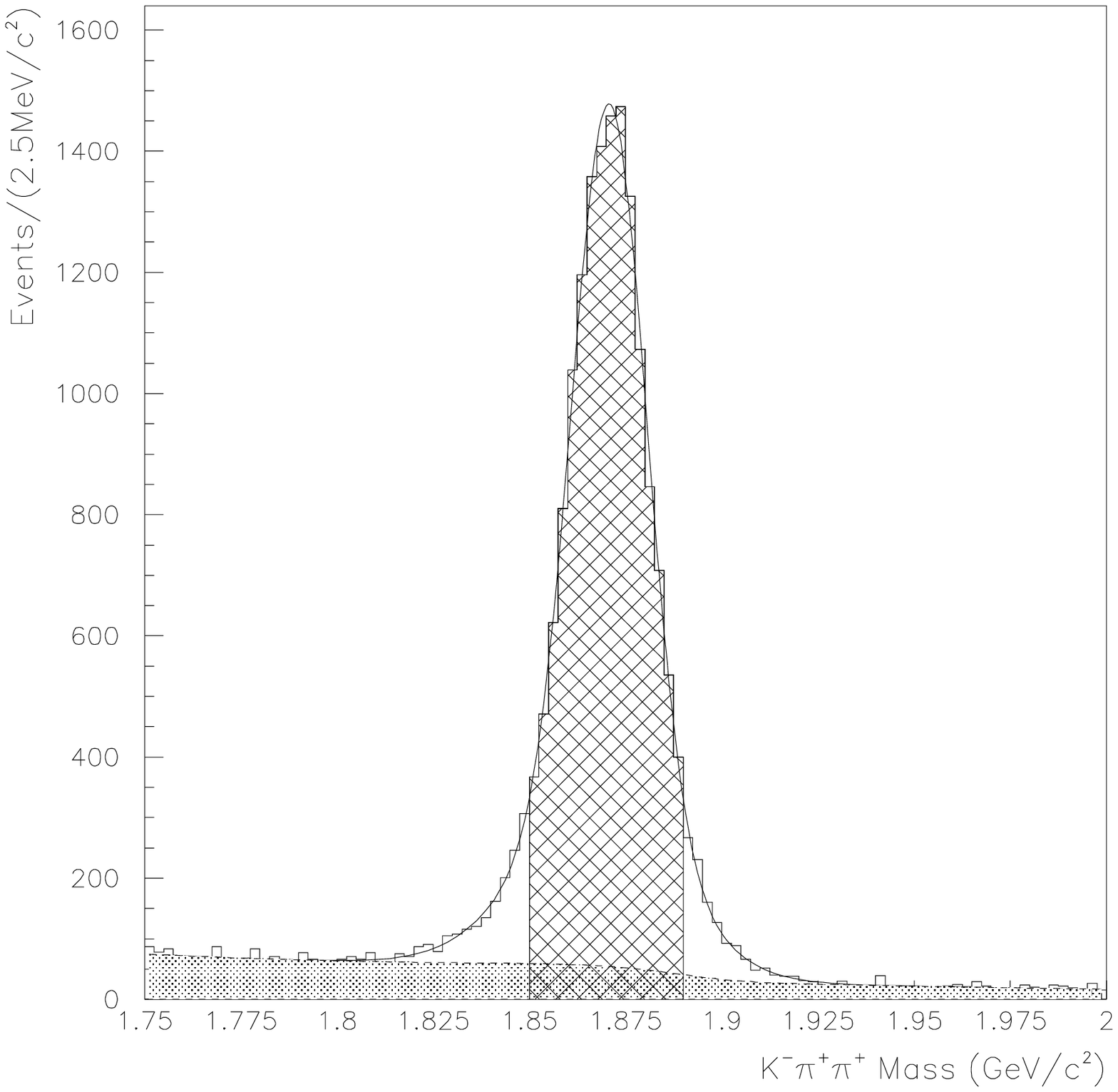}} 
\resizebox{8cm}{!}{\includegraphics{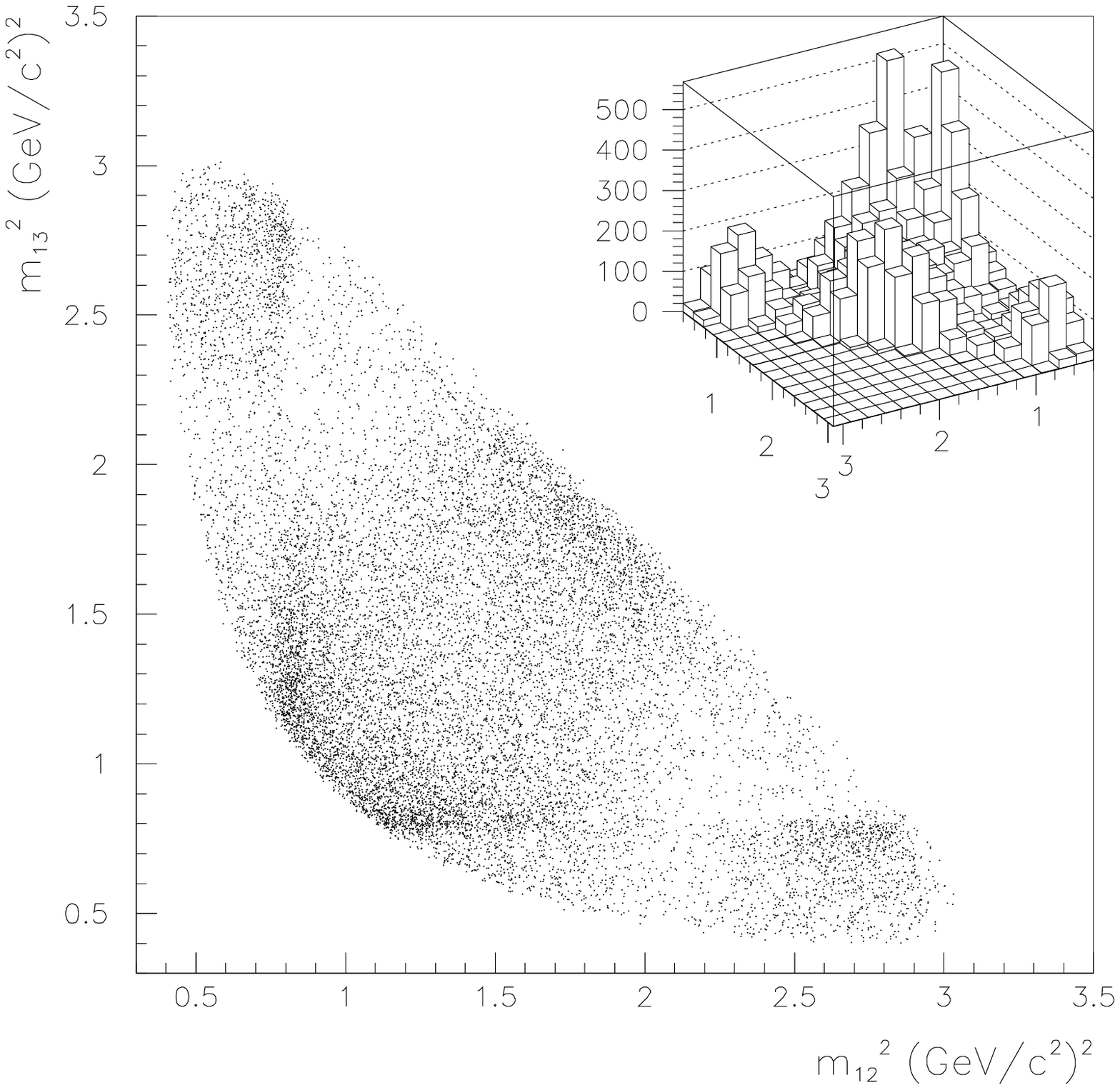}}
  \put(-400,180){(a)}\put(-170,180){(b)}
\caption{(a) The \kpp invariant mass spectrum. The filled area is 
background; (b) Dalitz plot corresponding to the events in the dashed area of (a).
\label{kpipi} }
\end{figure}

From the original $2\times 10^9$ events collected by E791, and after reconstruction and
selection criteria, we obtained the \kpipi sample shown in Figure~\ref{kpipi}(a). 
The filled area represents the level of background; besides the combinatorial, the other main source 
of background comes from the reflection of the decay $D_s^+\to K^-K^+\pi^+$ (through $\bar K^*K^+$ 
and $\phi\pi^+$). 
The crosshatched region contains the events selected for the Dalitz-plot ana\-lysis. 
There are 15090 events in this sample, of which 6\% are background. 

Figure~\ref{kpipi}(b) shows the Dalitz-plot for these events. The two axes are the squared 
invariant-mass combinations for $K\pi$, and the plot
is symmetrized with respect to the two identical pions.
The plot presents a rich structure, where we can observe the clear
bands from $\bar K^*(890)\pi^+$, and an accumulation of events at the upper edge of
the diagonal, due to heavier resonances.
To study the resonant substructure, we perform an unbinned maximun-likelihood fit to the 
data, with probability distribution funtions (PDF's) for both signal and background sources.
In particular, for each candidate event, the signal PDF is written as the square of the 
total physical amplitude ${\cal A}$ (defined below) and it is weighted
for the acceptance across the Dalitz plot (obtained by Monte Carlo (MC)) and by 
the level of signal to background for each event, as given by the line shape of 
Figure~\ref{kpipi}(a). The background PDF's (levels and shapes) are fixed for the Dalitz-plot fit, 
according to MC and data studies.

We begin describing our first approach to fit the data, which represents the conventional
Dalitz-plot analysis including the known $K\pi$ resonant amplitudes (${\cal A}_n,~n\ge 1$), plus 
a constant non-resonant contribution. The signal amplitude is constructed as a coherent sum of 
the various sub-channels:
\begin{equation}
{\cal A} = a_0e^{i\delta_0}{\cal A}_0 + \sum_{n=1}^N a_ne^{i\delta_n}
{\cal A}_n(m^2_{12},m^2_{13})
\label{ampl}
\end{equation} 
Each resonant amplitude  is written as 
\begin{equation}
{\cal A}_n = \ BW_n~F_D^{(J)}~ F_n^{(J)}~ {\cal M}_n^{(J)}~.
\label{ampl_n}
\end{equation}
where $BW_n$ is the relativistic Breit-Wigner propagator 
\begin{equation}
 BW_n = {1 \over {m^2_0 - m^2 - im_0\Gamma(m)}} 
\label{bw}
\end{equation} 
with mass dependent width,
\begin{equation}
\Gamma(m) = \Gamma_0 \frac{m_0}{m}\left(\frac{p^*}{p^*_0}\right)^{2J+1}
\left(\frac{F_n^{(J)}(p^*)}{F_n^{(J)}(p^*_0)}\right)^2\ .
\label{gamma}
\end{equation} 
The quantities $F_D$ and  $F_R$ are the Blatt-Weisskopf damping factors
respectively for the $D$ and the $K\pi$ resonances, they depend on the 
radii of the decaying meson and are set to $r_D=3.0$ GeV$^{-1}$ and $r_R=1.5$ GeV$^{-1}$ 
\cite{argus};
$p^*$ is the pion momentum in the resonance rest frame at mass $m_{12}\
(p^*_0=p^*(m_0))$. ${\cal M}_n^{(J)}$ describes the angular distribution due to the 
spin $J$ of the resonance. See details in \cite{d3pi}. Finally each amplitude is Bose
symmetrized ${\cal A}_n = {\cal A}_n[({\bf 12}){\bf3}] + 
{\cal A}_n[({\bf 13}){\bf 2}]$.

Using this model (Model A), we find contributions from the following channels: the non-resonant,
responsible for more than 90\% of the decay rate, followed by $\bar K^*_0(1430)\pi^+$, 
$\bar K^*(892)\pi^+$, $\bar K^*(1680)\pi^+$ and $\bar K^*_2(1430)\pi^+$. The decay fractions 
and relative phases are shown in Table~\ref{tablekappa}. 
These values are in accordance with previous results from E691 \cite{e691-kpipi} and E687
\cite{e687-kpipi}. We thus confirm a high
non-resonant contribution according to this model, which is totally unusual in $D$ decays.
Besides, there is an important destructive interference pattern, since all fractions add up
to 140 \%.

\begin{figure}[t]
\resizebox{8cm}{!}{\includegraphics{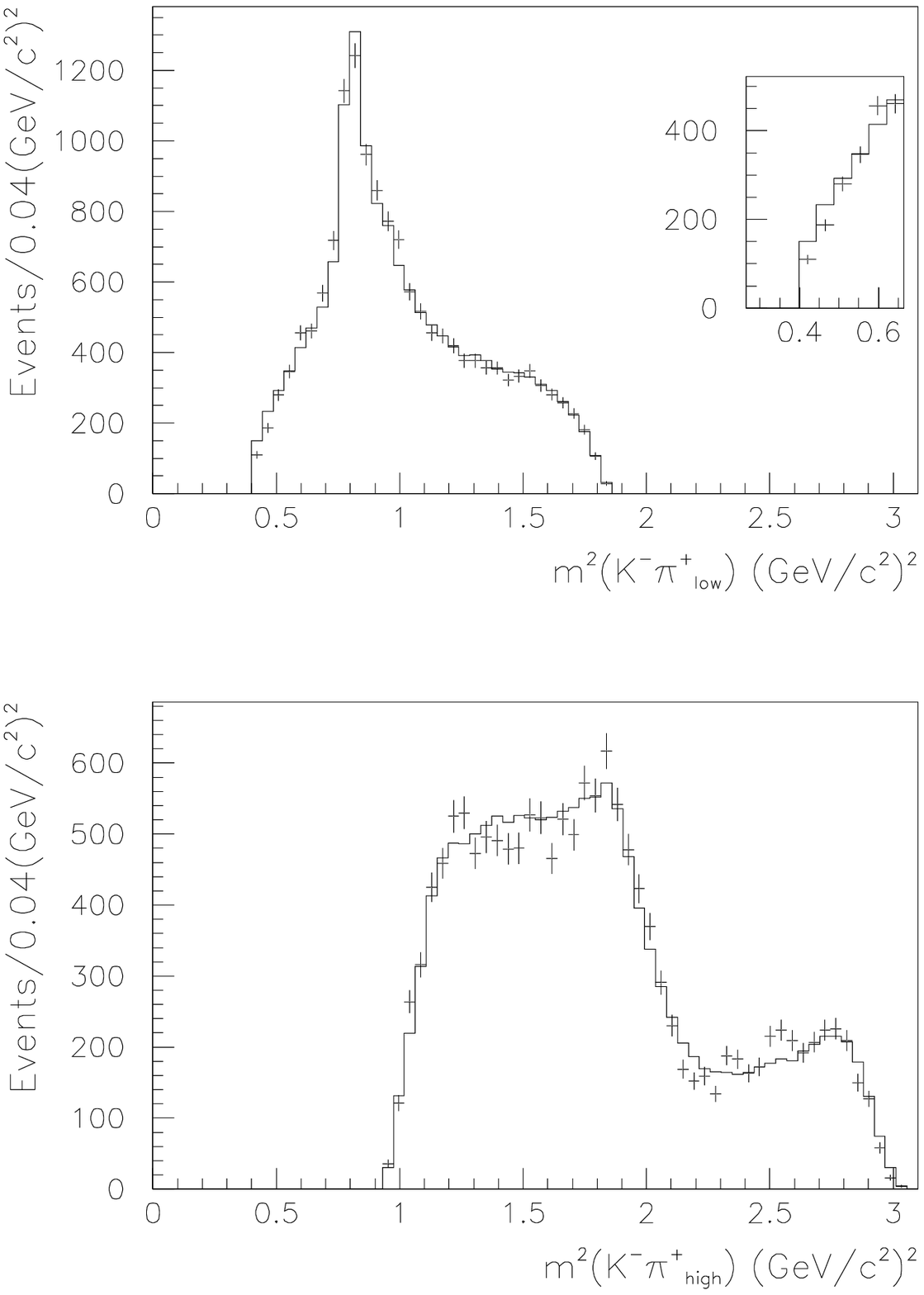}} 
\resizebox{8cm}{!}{\includegraphics{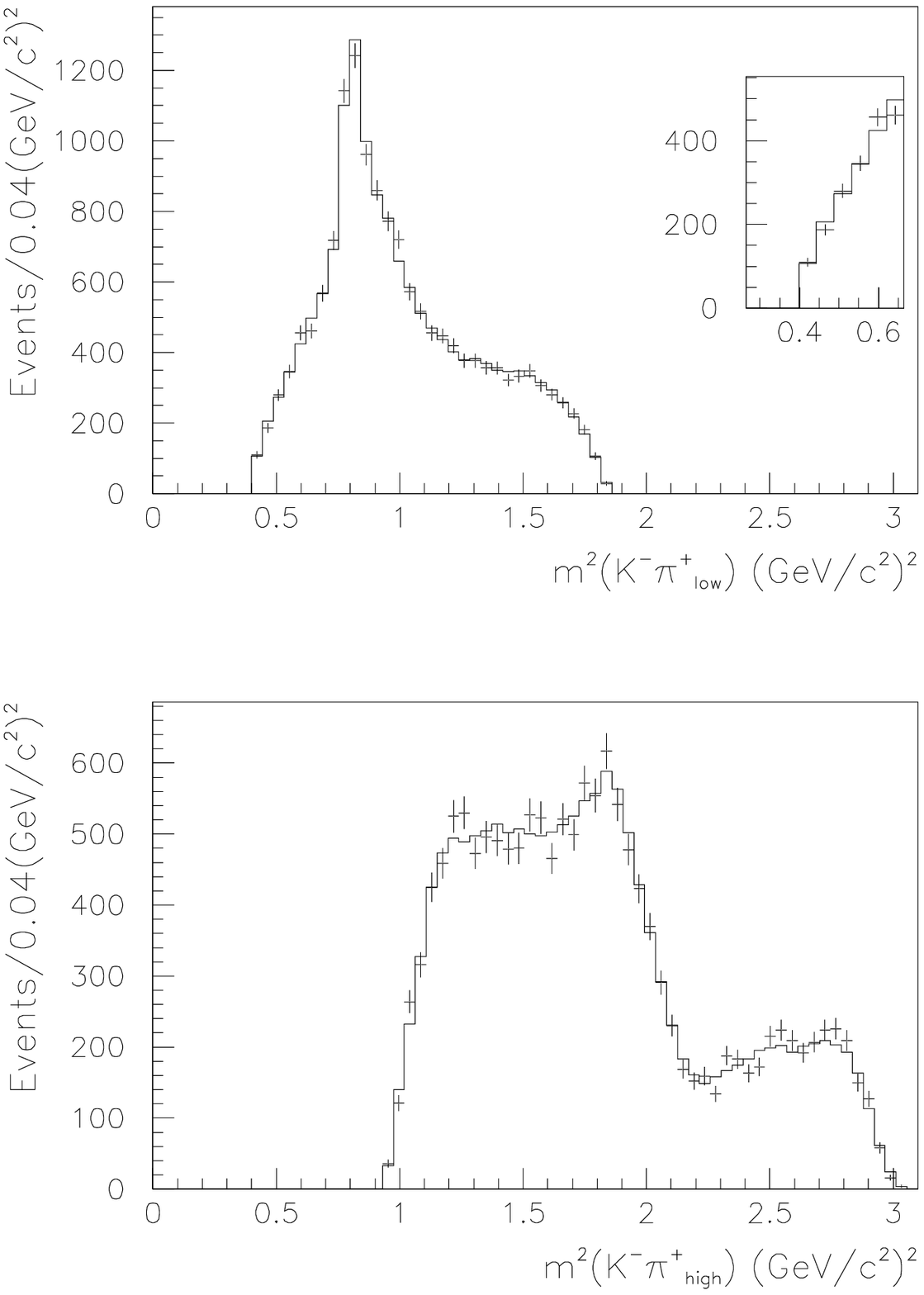}} 
  \put(-350,-5){(a)}\put(-120,-5){(b)}
\caption{$m^2(K\pi_{\rm low})$ and $m^2(K\pi_{\rm high})$ projections for data
(error bars) and fast MC (solid line): (a) fit to Model A, without $\kappa$, and (b) fit to Model B,
with $\kappa$.}
\label{proj_kpipi} 
\end{figure}

To evaluate the fit quality of Model A, we compute a two-dimensional 
$\chi^2$  in the Dalitz plot, from the difference 
in densities for the model (from a fast-MC algorithm) and the data. 
We obtain $\chi^2/\nu=2.7$ ($\nu$ being the number of degrees of freedom), with a corresponding 
confidence level (CL) of $10^{-11}$. In Figure~\ref{proj_kpipi}(a) we show the $K\pi$ low and
high squared-mass projections for data (error bars) and model (solid line). The discrepancies
are evident at the very low-mass region for $m^2(K\pi)_{\it low}$ and near
2.5 (GeV/c$^2$)$^2$ for  $m^2(K\pi)_{\it high}$. These regions of disagreement are the 
same observed previously by E687 \cite{e687-kpipi}. We thus conclude that a model with 
the known $K\pi$ resonances, plus a non-resonant amplitude, is not able to describe the 
\kpipi Dalitz plot satisfactorily.

A similar pattern -- bad fit quality with large NR fraction -- is found in the analysis of the
decay \d3pi when allowing only the established $\pi\pi$ resonances \cite{d3pi} .
There we find that the inclusion of an extra scalar resonance improves the fit substantially, giving
strong evidence for the $\sigma(500)$. See the section on \d3pi below.
Thus, we are lead to try an extra scalar resonance in our fit model here.
The possible existence of a light and broad $K\pi$ scalar state has been suggested
by many authors \cite{beveren,ishida1,black1,pelaez,shakin}, some of them believing it would be
a member of a light scalar nonet \cite{ishida2,black2,oller}; however, its existence 
has been the subject of some controversy also\cite{torn,aniso,cherry}. 

A second fit model, Model B, is constructed by the inclusion of an extra scalar state, 
with unconstrained mass
and width. For consistency, the mass and width of the other scalar state, the $K^*_0(1430)$, are 
also free parameters of the fit. We adopt a better description for these scalar states by 
introducing gaussian-type form-factors \cite{torn} to take into account the finite size of the
decaying mesons. Two extra floating parameters are the meson radii $r_D$ and $r_R$ introduced
above.

Using this model, we obtain the values of  $797\pm 19\pm 42$ MeV/c$^2$ for the mass and 
$410\pm 43\pm 85$ MeV/c$^2$ for the width of the new scalar state (first error statistical, 
second error systematic), referred to here as the $\kappa$.
The values of mass and width obtained for the $K^*_0(1430)$ are respectively 
$1459\pm 7\pm 6$ MeV/c$^2$ and $175\pm 12\pm 12$ MeV/c$^2$, appearing heavier and
narrower than presented by the PDG \cite{pdg}. The decay fractions and relative phases for 
Model B, with systematic errors, are given in Table~\ref{tablekappa}. 
Compared to the results of Model A (without $\kappa$), 
the non-resonant mode drops from over 90\% to 13\%. The $\kappa\pi^+$ state 
is now the dominant channel with about 50\%. 
The meson radii $r_D$ and $r_R$ are found to be respectively $5.0\pm 0.5$ GeV$^{-1}$ and 
$1.6\pm 1.3$ GeV$^{-1}$, in complete agreement with previous estimates \cite{argus,cleo}.

\begin{table}\centering
\begin{tabular}{|c|c c|c c|} \hline 
 Decay & \multicolumn{2}{|c|}{Model A: No $\kappa$} 
 & \multicolumn{2}{|c|}{Model B: With $\kappa$ }  \\
 Mode  & Fraction (\%) & Phase & Fraction (\%) & Phase  \\  \hline
NR           & $90.9\pm 2.6$ & $0^\circ$ (fixed) & $13.0\pm 5.8\pm 2.6$ & $(349\pm 14\pm 8)^\circ$  \\ 
$\kappa\pi^+$  & -- & -- & $47.8\pm 12.1\pm 3.7$ & $(187\pm 8\pm 17)^\circ$ \\ 
$\bar K^*(892)\pi^+$  & $13.8\pm 0.5$ & $(54\pm 2)^\circ$ & $12.3\pm 1.0\pm 0.9$ & $0^\circ$ (fixed)  \\ 
$\bar K^*_0(1430)\pi^+$ & $30.6\pm 1.6$ & $(54\pm 2)^\circ$ & $12.5\pm 1.4\pm 0.4$ & $(48\pm 7\pm 10)^\circ$ \\ 
$\bar K^*_2(1430)\pi^+$   & $0.4\pm 0.1$ & $(33\pm 8)^\circ$ & $0.5\pm 0.1\pm 0.2$ & $(306\pm 8\pm 6)^\circ$\\ 
$\bar K^*(1680)\pi^+$    & $3.2\pm 0.3$ & $(66\pm 3)^\circ$ & $2.5\pm 0.7\pm 0.2$ & $(28\pm 13\pm 15)^\circ$\\ \hline 
\end{tabular}
\caption{Results without $\kappa$ (Model A) and with $\kappa$ (Model B). {\it Preliminary}.
\label{tablekappa}}
\end{table}
Moreover, the fit quality of Model B is substantially superior to that of Model A. 
The $\chi^2/\nu$ is now 0.73 with a CL of 95\%. The very good agreement between
the model and the data can be seen in the projections of Figure~\ref{proj_kpipi}(b).

A number of studies were done to check these results. For example, we replaced the 
complex $\kappa$ Breit-Wigner by a real Breit-Wigner, with no phase variation. In this case, we got
similar mass and width for this extra state, but with unphysical fractions for this state and
the NR, and a worse fit quality. We also replaced the $\kappa$ by a hypothetical vector state, 
with unconstrained mass and width, but it appears with a small fraction, the fit quality being 
comparable to the model without it, and all fractions and phases remaining unchanged. A tensor model 
was also tried without convergence, the width being driven to large, negative values.
Other models with the $\kappa$ were also tried. For example, modifications 
to the scalar Breit-Wigner amplitude and to the form-factors were introduced. 
A number of studies for the parameterization of the NR amplitude were tried \cite{bediaga}, 
with and without the $\kappa$. No model without the $\kappa$ was able to describe our data
satisfactorily. All variations of models with $\kappa$ gave similar results for the $\kappa$ 
mass and width (within errors) although the fractions for $\kappa\pi$ and NR showed correlations.

Thus, from the results above, we find strong evidence that a light and broad scalar $K\pi$
resonance gives an important contribution to the \kpipi decay.

\section{The \ds3pi Results}

\begin{figure}[t]
\resizebox{8cm}{!}{\includegraphics{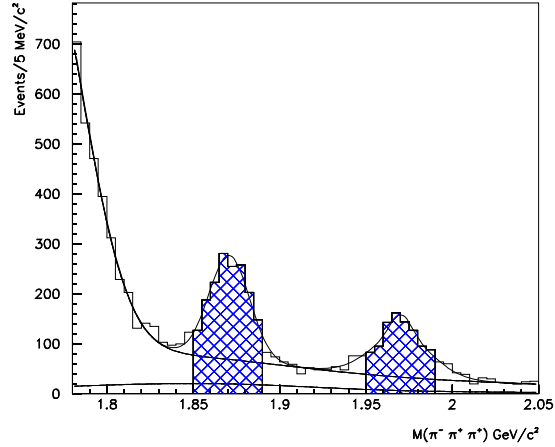}} 
\caption{The $\pi^-\pi^+\pi^+$ invariant mass spectrum. The dotted line
represents the $D^0\to K^-\pi^+$ plus $D_s^+\to \eta'\pi^+$ reflections and 
the dashed line is the total background. Events used for the Dalitz analyses
are in the hatched areas.}
\label{m3pi} 
\end{figure}

\begin{figure}[t]
\resizebox{7cm}{!}{\includegraphics{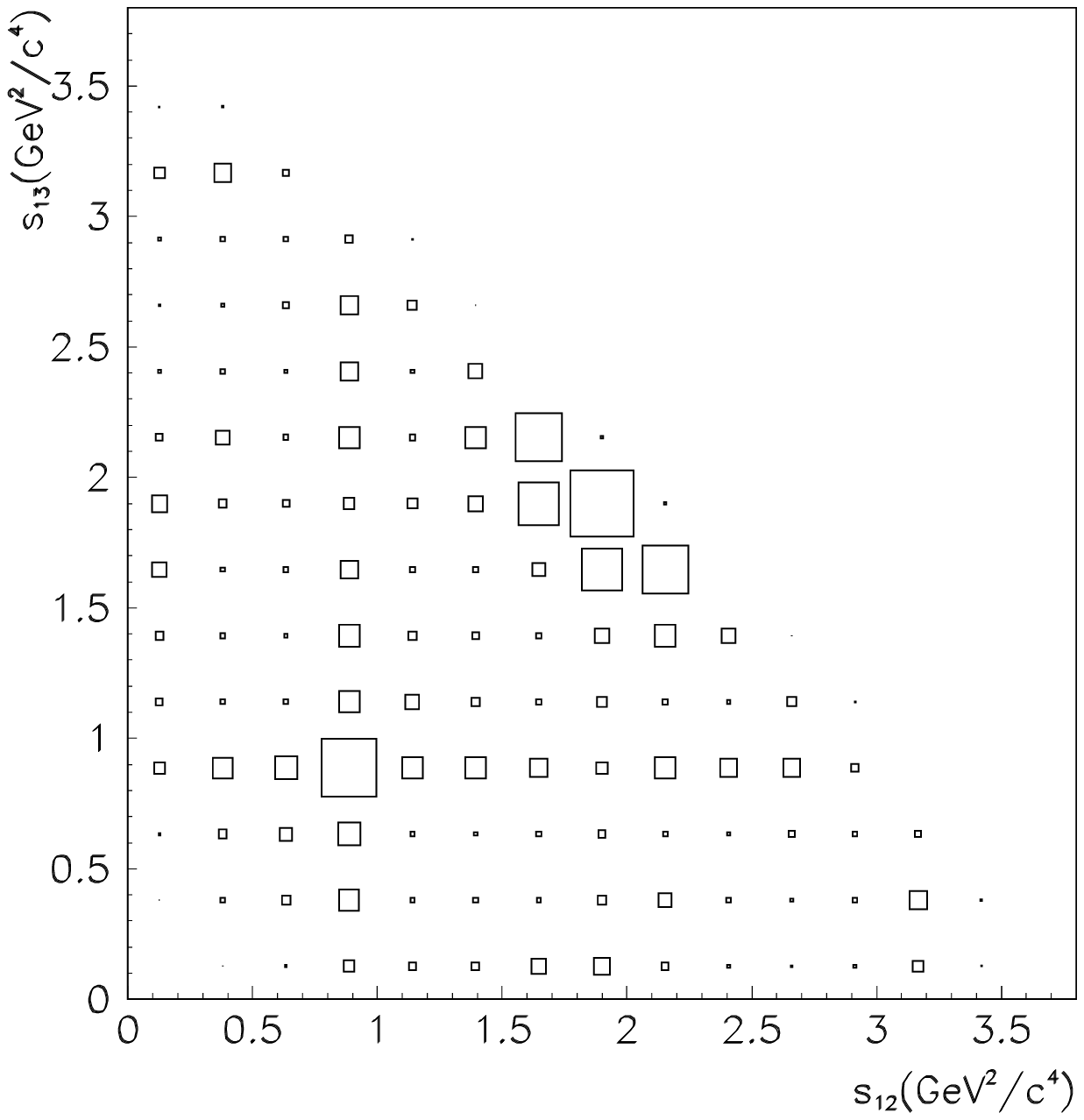}} 
\resizebox{7.3cm}{!}{\includegraphics{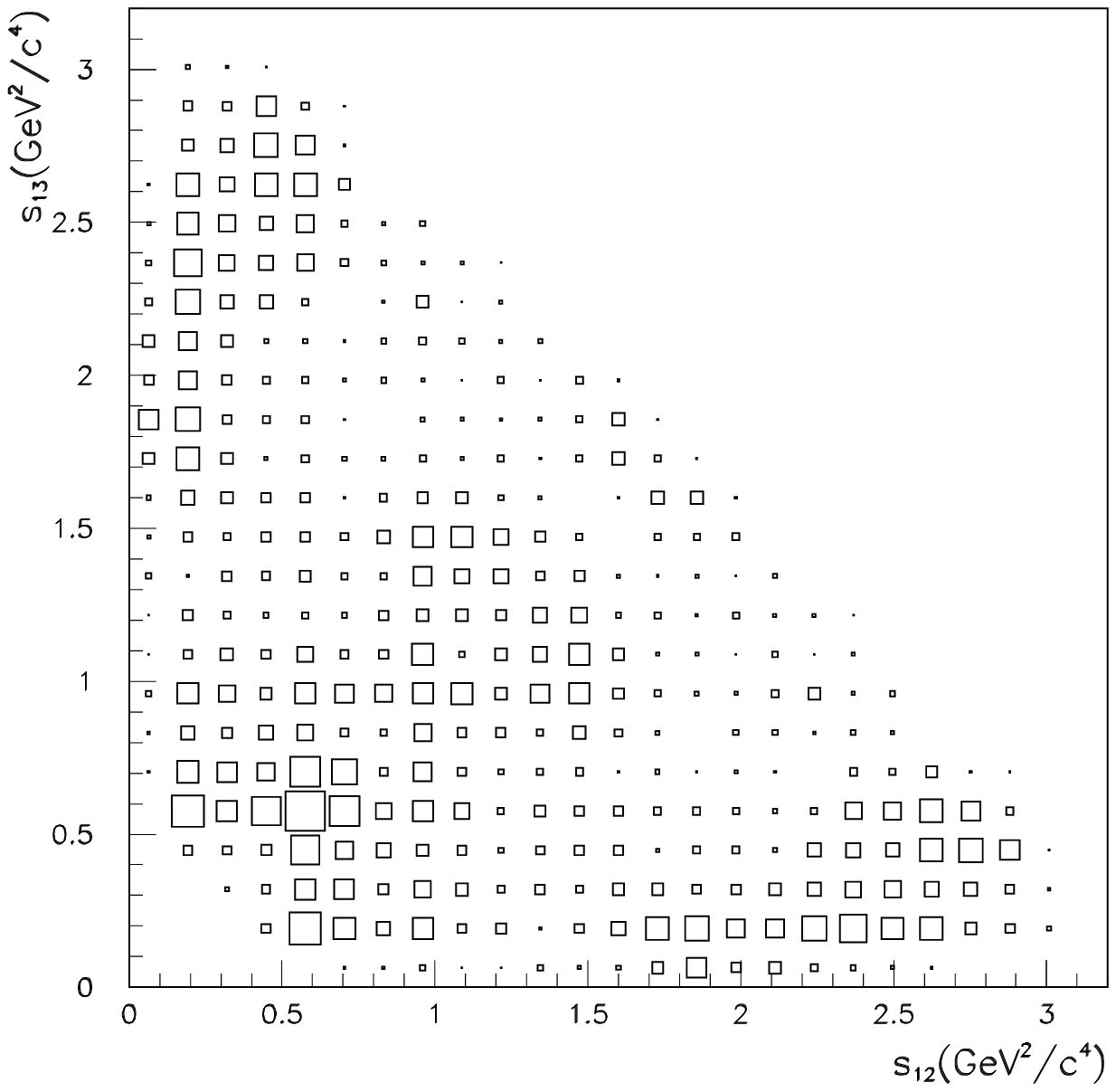}}
  \put(-300,160){(a)}\put(-70,160){(b)}
\caption{(a) The $D_s^+ \to \pi^- \pi^+ \pi^+$ Dalitz plot and
(b) the \d3pi Dalitz plot. Since there are two identical pions, 
the plots are symmetrized.
\label{dalitz3pi} }
\end{figure}

In Figure~\ref{m3pi} we show the $\pi^-\pi^+\pi^+$ invariant mass distribution for
the sample collected by E791 after reconstruction and selection criteria \cite{ds3pi,d3pi}. 
Besides combinatorial background, reflections from the decays \kpipi, $D^+\to K^-\pi^+$ 
(plus one extra track) and $D_s^+\to \eta'\pi^+,
~\eta'\to\rho^0(770)\gamma$ are all taken into account.
The hatched regions in Figure \ref{m3pi} show the samples used for the
Dalitz-plot analyses. There are 1686 and 937 candidate events for $D^+$  and $D_s^+$respectively,
with a signal to background ratio of about 2:1. The Dalitz plots for these events are shown 
in Figure~\ref{dalitz3pi}, the axes corresponding to the two $\pi^-\pi^+$ invariant-masses 
squared.
For the Dalitz-plot fits of both \d3pi and \ds3pi decays, we use essentially the same formalism
as for the \kpipi decays. See details in \cite{ds3pi,d3pi}.

For the \ds3pi events in Figure~\ref{dalitz3pi}(a), the signal amplitude includes
the following channels: $\rho^0(770) \pi^+$, 
$f_0(980) \pi^+$, $f_2(1270) \pi^+$, $f_0(1370) \pi^+$, $\rho^0(1450) \pi^+$ and the 
non-resonant, assumed constant
across the Dalitz plot. 

For the  $f_0(980)\pi^+$ amplitude, instead of a simple Breit-Wigner of Eq.~\ref{bw}
\footnote{For both $D,D_s\to 3\pi$ analyses, the relativistic 
Breit-Wigner for each resonant amplitude is defined with a factor (-1) with respect to 
Eq.~\ref{bw}.},
we use a coupled-channel Breit-Wigner function \cite{wa76},
\begin{equation}
BW_{f_0(980)} = {1 \over {m_{\pi\pi}^2 - m^2_0 + im_0(\Gamma_{\pi}+\Gamma_K)}}\ ,
\end{equation}
\begin{equation}
\Gamma_{\pi} = g_{\pi}\sqrt{m_{\pi\pi}^2/ 4 - m_{\pi}^2},~
\Gamma_K = {g_K \over 2} \left( \sqrt{m_{\pi\pi}^2/ 4 - m_{K^+}^2}+
\sqrt{m_{\pi\pi}^2/ 4 - m_{K^0}^2}\right) .
\end{equation}
The \ds3pi Dalitz plot is fit to obtain not only the 
decay fractions and phases of the possible sub-channels, but also 
the parameters of the $f_0(980)$ state, $g_{\pi}$, $g_K$, and $m_0$, as well 
as the mass and width of the $f_0(1370)$. The other resonance 
masses and widths are taken from the PDG\cite{pdg}.
The resulting fractions and phases are given in Table \ref{tabds3pi}. 

The measured $f_0(980)$ parameters are 
 $m_0 =  977 \pm 3 \pm 2$ MeV/c$^2$, $g_{\pi} =$ 0.09  $\pm$  0.01 $\pm$ 0.01 
and  $g_K =$ 0.02  $\pm$  0.04  $\pm$  0.03.
Our value for $g_{\pi}$ is in very good agreement with OPAL and MARKII results \cite{opal}, 
but WA76 \cite{wa76} found a much larger value,
$g_{\pi} =$ 0.28  $\pm$  0.04. Our value of $g_K$ indicates a small coupling of $f_0(980)$ to
$K\bar K$. The values of the $f_0(980)$ mass and of
$g_{\pi}$, as well as  the magnitudes and phases of the resonant amplitudes, 
are relatively insensitive to the value of $g_K$. Both OPAL and MARKII results
are also insensitive to the value of $g_K$. WA76, on the contrary, measured 
$g_K =$ 0.56  $\pm$  0.18. 

By fitting the Dalitz plot using for the $f_0(980)$ a simple Breit-Wigner function, we find 
$m_0 = 975 \pm 3$ MeV/c$^2$ and $\Gamma_0 =  44 \pm 2 \pm 2$ MeV/c$^2$, and the results
for fractions and phases are indistinguishable.

\begin{figure}[t]
\resizebox{10cm}{!}{\includegraphics{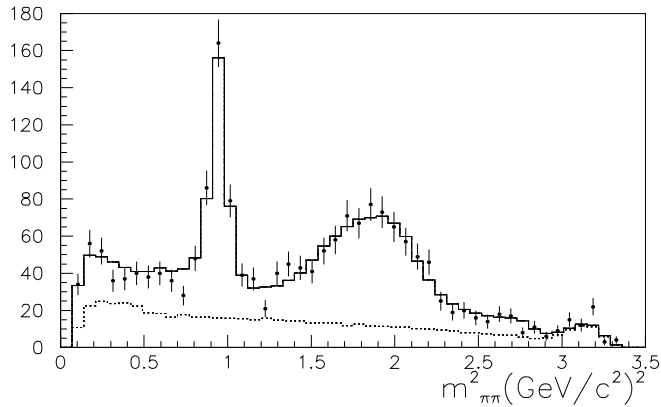}}
\caption{$s_{12}$ and $s_{13}$ ($m^2_{\pi\pi}$) projections for 
\ds3pi data (dots) and our best fit
(solid). The hashed area corresponds to  background.
\label{proj_ds} }
\end{figure}

The confidence level of the fit for \ds3pi is 35\% \cite{ds3pi}. In Figure~\ref{proj_ds} we show
the $\pi^-\pi^+$ mass-squared projections for data (points)
and model (solid lines, from fast-MC).

\begin{table}[t]\centering
\begin{tabular}{|c|c c|} \hline
Decay Mode          &    Fraction (\%)              &  Phase   \\ \hline 
$f_0(980)\pi^+$     & $56.5\pm 4.3\pm 4.7$ & $0^\circ$ (fixed)   \\ 
NR          & $ 0.5\pm 1.4\pm 1.7$ &~$(181\pm 94\pm 51)^\circ$~\\ 
$\rho^0(770)\pi^+$  & $ 5.8\pm 2.3\pm 3.7$ & $(109\pm 24\pm  5)^\circ$ \\ 
$f_2(1270)\pi^+$    & $19.7\pm 3.3\pm 0.6$ & $(133\pm 13\pm 28)^\circ$ \\ 
$f_0(1370)\pi^+$    & $32.4\pm 7.7\pm 1.9$  & $(198\pm 19\pm 27)^\circ$ \\ 
$\rho^0(1450)\pi^+$ & $ 4.4\pm 2.1\pm 0.2$ & $(162\pm 26\pm 17)^\circ$ \\ \hline
\end{tabular}
\caption{Dalitz fit results for \ds3pi.}
\label{tabds3pi}
\end{table}

As we can see by the results of Table \ref{tabds3pi}, approximately half of 
the $D_s^+ \to \pi^- \pi^+ \pi^+$ 
rate is via  $f_0(980)\pi^+$. If the spectator amplitude is dominant in 
this decay, this would support the interpretation of the $ f_0(980) $
as an $ s \bar s $ state.  On the other hand, the 
large contribution from $ f_0 (1370) \pi^+$ 
indicates the  presence of either $ W$-annihilation amplitudes or strong 
rescattering in the final state. In fact, the $ f_0 (1370) \pi^+$ is not observed
in the $D_s^+ \to K^+K^-\pi^+$  final state\cite{e687-kkpi}, pointing to the
$f_0(1370)$ being  a  non-$ s \bar s $ state, as suggested by the 
naive quark model\cite{pdg}. There is no evidence in the $D_s^+$ decay for a low-mass
broad scalar particle as seen in the  $D^+\to \pi^- \pi^+ \pi^+$ decay, discussed below.

\section{The \d3pi Decay}
\label{secd3pi}

\begin{figure}[t]
\resizebox{8cm}{!}{\includegraphics{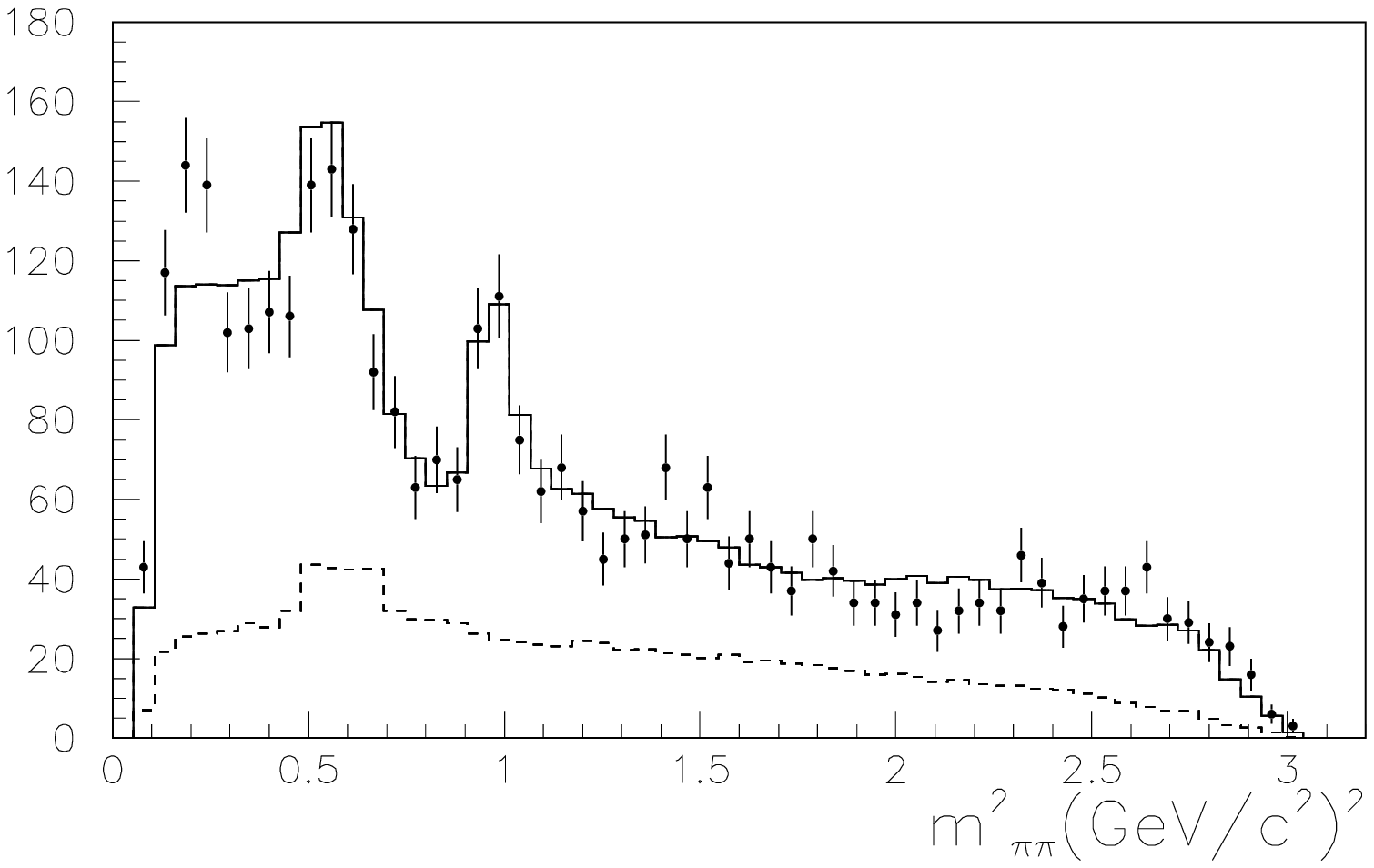}} 
\resizebox{8cm}{!}{\includegraphics{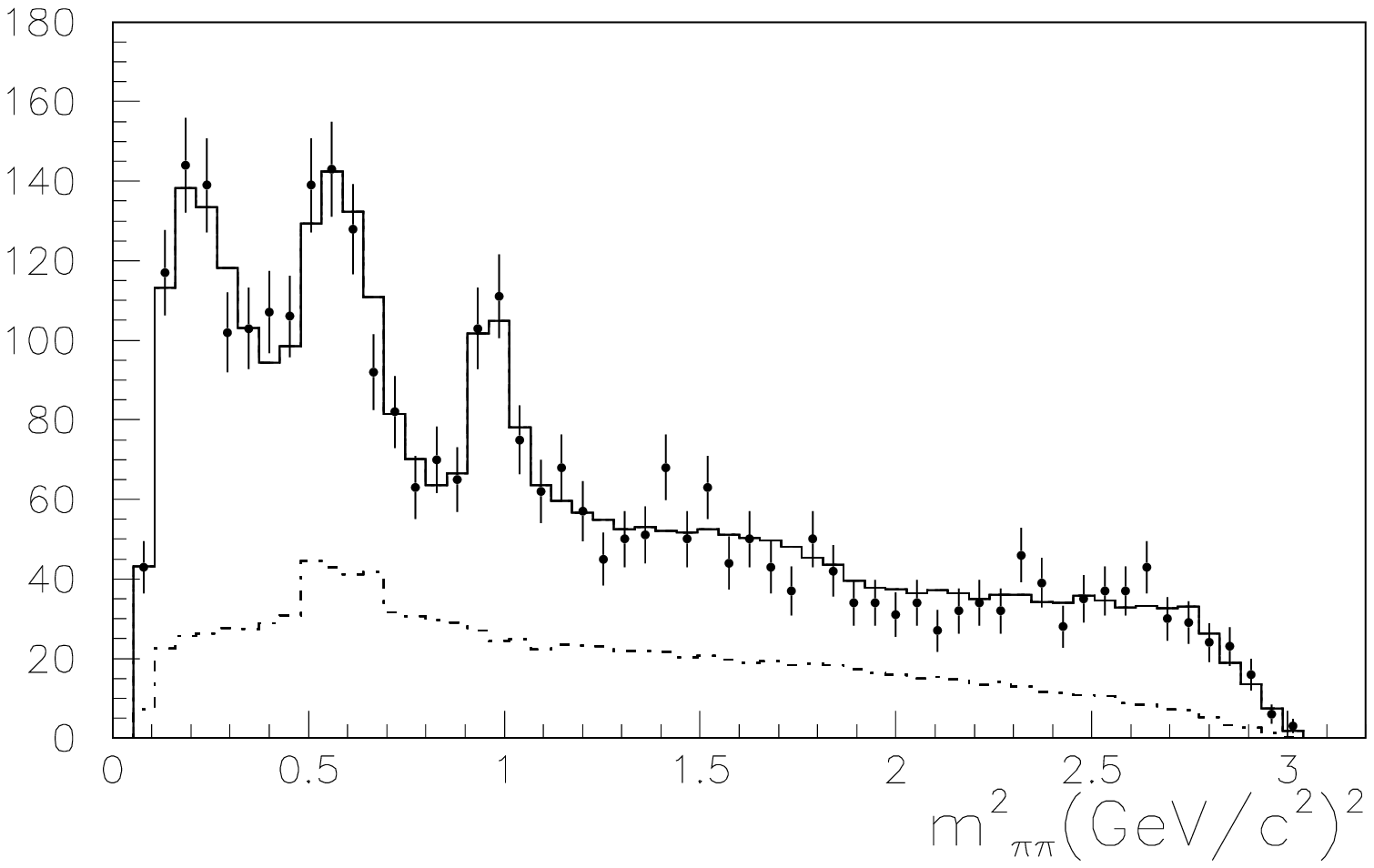}} 
  \put(-300,100){(a)}\put(-70,100){(b)}
\caption{$s_{12}$ and $s_{13}$ ($m^2_{\pi\pi}$) projections for \d3pi data (dots) 
and our best fit (solid) for models \rm (a) without and \rm (b) with $\sigma\pi^+$ amplitude. 
The dashed distribution corresponds to the expected background level.}
\label{proj_dp3pi} 
\end{figure}

In a first approach, we try to fit the \d3pi Dalitz plot of Figure~\ref{dalitz3pi}(b)
with the same amplitudes used for the \ds3pi analysis. Using this model, 
the non-resonant, the $\rho^0(1450)\pi^+$, and the $\rho^0(770)\pi^+$ amplitudes
are found to dominate, as shown in  Table~\ref{tabd3pi}, and in agreement with
previous reported analyses \cite{e691-3pi,e687-3pi}.
However, this model does not describe the data satisfactorily, especially at low 
$\pi^-\pi^+$ mass squared, as can be seen from Fig.~\ref{proj_dp3pi}(a). The $\chi^2/\nu$ 
obtained from the binned Dalitz plot for this model is 1.6, with a CL less 
than $10^{-5}$. 
   
To investigate the possibility that another $\pi^-\pi^+$ resonance contributes to the
\d3pi decay, we add an extra scalar resonance amplitude to the signal PDF, with  
mass and width as floating parameters in the fit.

We find that this model improves our fit substantially. 
The mass and the width of the extra scalar state are found to be 
$ 478^{+24}_{-23} \pm 17  $ MeV/$c^2$ and $ 324^{+42}_{-40}  \pm 21$ MeV/$c^2$, respectively.
Refering to this state as the $\sigma$, we obtain that the
$\sigma\pi^+$ chanell produces the largest decay fraction, as shown in Table~\ref{tabd3pi}; 
the non-resonant amplitude, which is dominant in the model without 
$\sigma\pi^+$, drops substantially. 
This model describes the data much better, as can be seen by the $\pi\pi$ mass squared
projection in Fig.~\ref{proj_dp3pi}(b). The $\chi^2/\nu$ is now 0.9, with a corresponding  
confidence level of 91\%.

The existence of a light $\pi\pi$ state, or the $\sigma$, has been the subject of a 
long-standing controversy \cite{penni2,torn3}. Various experiments have presented
inconsistent evidence for this state \cite{sigma-exp}, yielding conflicting results 
\cite{pdg,torn3}. 

To test the model above, we replace the scalar amplitude by
vector and tensor states, and also by a real Breit-Wigner, with no phase variation (as 
also done in the \kpipi analysis).
All these alternative models fail to describe the data as well as the scalar (regular)
Breit-Wigner amplitude. See detailed discussion in \cite{d3pi}.

\begin{table}[t]\centering
\begin{tabular}{|c|c c|c c|} \hline  
Decay & \multicolumn{2}{|c|}{Fit without $\sigma\pi^+$} 
      & \multicolumn{2}{|c|}{Fit with $\sigma\pi^+$} \\ 
Mode               &    Fraction (\%) &  Phase
                   &    Fraction (\%)  &    Phase  \\ \hline 
$\sigma\pi^+$      &         --               &          -- 
                   & $46.3\pm 9.0\pm 2.1$ & $(206\pm 8\pm 5)^\circ$\\ 
$\rho^0(770)\pi^+$ & $20.8\pm 2.4$  & $0^\circ$ (fixed) 
                   & $33.6\pm 3.2\pm 2.2$ & $0^\circ$ (fixed)     \\ 
NR         & $38.6\pm 9.7$  & $(150\pm 12)^\circ$     
                   & $ 7.8\pm 6.0\pm 2.7$ & $(57\pm 20\pm 6)^\circ$  \\ 
$f_0(980)\pi^+$    & $7.4\pm 1.4$   & $(152\pm 16)^\circ$    
                   & $ 6.2\pm 1.3\pm 0.4$ & $(165\pm 11\pm 3)^\circ$ \\ 
$f_2(1270)\pi^+$   & $6.3\pm 1.9$   & $(103\pm 16)^\circ$   
                   & $19.4\pm 2.5\pm 0.4$ & $(57\pm 8\pm 3)^\circ$  \\ 
$f_0(1370)\pi^+$   & $10.7\pm 3.1$  & $(143\pm 10)^\circ$     
                   & $ 2.3\pm 1.5\pm 0.8$ & $(105\pm 18\pm 1)^\circ$ \\ 
$\rho^0(1450)\pi^+$& $22.6\pm 3.7$  & $ (46\pm 15)^\circ$    
                   & $ 0.7\pm 0.7\pm 0.3$ & $(319\pm 39\pm 11)^\circ$\\ \hline
\end{tabular}
\caption{Dalitz fit results for \d3pi. First errors are statistical, second 
systematics (only for fit with $\sigma\pi^+$ mode).}
\label{tabd3pi}
\end{table}
\section{Conclusion}

From the data of the Fermilab E791 experiment, we studied the Dalitz plots of the
decays $D^+\to K^-\pi^+\pi^+$, \ds3pi and $D^+\rightarrow \pi^-\pi^+\pi^+$. 
In these three final states, the scalar intermediate resonances were found to give the main
contribution to the decay rates. We obtained strong evidence for the existence of the 
$\sigma$ and $\kappa$ scalar mesons, 
measuring their masses and widths. We also obtained new measurements for masses and
widths of the other scalars studied, $f_0(980)$, $f_0(1430)$ and $K^*_0(1430)$.

The results presented here show the potential of $D$ meson decays for the study of 
light meson espectroscopy, in particular in the scalar sector.


\begin{thebibliography}{99.}

\bibitem{ds3pi} E791 Collaboration, E.M.~Aitala {\em{et al.}}, Phys. Rev. Lett. {\bf 86}
765 (2001).
\bibitem{d3pi} E791 Collaboration, E.M.~Aitala {\em{et al.}}, Phys. Rev. Lett. {\bf 86}
770 (2001).
\bibitem{ref791} J.A.~Appel, Ann. Rev. Nucl. Part. Sci.  {\bf 42}, 367 (1992); D. Sum\-mers 
{\em et al.}, hep-ex/0009015; S. Amato {\em et al.}, Nucl. Instr. Meth. A {\bf 324}, 535 (1993);
E791 Collaboration, E.M. Aitala {\em et al.}, Eur. Phys. J. direct C {\bf 4}, 1 (1999).
\bibitem{beveren} E.~van~Beveren {\em et al.}, Z. Phys. C {\bf 30}, 615 (1986).
\bibitem{ishida1} S.~Ishida {\em et al.}, Prog. Theor. Phys. {\bf 98}, 621 (1997).
\bibitem{black1} D.~Black {\em et al.}, Phys. Rev. D {\bf 58}, 054012 (1998).
\bibitem{pelaez} J.A.~Oller, E.~Oset, J.R.~Pel\'aez, Phys. Rev. D {\bf 59}
074001 (1999);  M.~Jamin, J.A.~Oller, and A.~Pich, Nucl. Phys. {\bf B587}, 331 (2000).
\bibitem{shakin} C.M.~Shakin, H.~Wang, Phys. Rev. D {\bf 63}, 014019 (2001).
\bibitem{ishida2} M.~Ishida, Prog. Theor. Phys. {\bf 101}, 661 (1999).
\bibitem{black2} D.~Black {\it et al.}, Phys. Rev. D {\bf 59}, 074026 (1999).
\bibitem{oller} J.A.~Oller and E.~Oset, Phys. Rev. D {\bf 60}, 074023 (1999).
\bibitem{torn} N.A.~T\"ornqvist, Z. Phys. C {\bf 68}, 647 (1995).
\bibitem{aniso} A.V.~Anisovitch and A.V.~Sarantsev, Phys. Lett. B {\bf 413}, 137 (1997).
\bibitem{cherry} S.N.~Cherry and M.R.~Pennington, Nucl. Phys. {\bf A688} 823 (2001).
\bibitem{e691-kpipi} E691 Collaboration, J.C.~Anjos {\em et al.}, Phys. Rev.
D {\bf 48}, 56 (1993).
\bibitem{e687-kpipi} E687 Collaboration, P.L.~Frabetti {\em et al.}, 
Phys. Lett. B {\bf 331}, 217 (1994).

\bibitem{pdg} Particle Data Group, D.E.~Groom {\em et al.}, Eur. Phys. Jour.
C {\bf 15}, 1 (2000).
\bibitem{argus} ARGUS Collaboration, H. Albrecht  {\em{et al.}}, Phys. Lett. B
{\bf 308}, 435(1993).
\bibitem{cleo} CLEO Collaboration, S.~Kopp {\em et al.}, Phys. Rev. D {\bf 63}, 
092001 (2001). 
\bibitem{blatt} J.M.~Blatt and V.F.~Weisskopf, Theoretical Nuclear Physics,
John Wiley \& Sons, New York, 1952.
\bibitem{lass} LASS Collaboration, D.~Aston {\em et al.}, Nucl. Phys. {\bf B296}, 
493 (1988).
\bibitem{bediaga} I.~Bediaga, C.~G\"obel, and R.~M\'endez-Galain, Phys. Rev.
Lett. {\bf 78}, 22 (1997) and Phys. Rev. D {\bf 56}, 4268 (1997); 
C.~G\"obel, Ph.D. Thesis, Centro Brasileiro de Pesquisas F\'\i sicas,
Rio de Janeiro, Brazil (1999).
\bibitem{wa76} WA76 Collaboration, T.A. Armstrong {\em et al.}, Z. Phys. C {\bf 51}, 
351 (1991). 
\bibitem{opal} OPAL Collaboration, K. Ackerstaff {\em et al.}, Eur. Phys. J. C {\bf 4}, 
19 (1998); MARKII Collaboration, G.~Gidal {\em et al.}, Phys. Lett. B {\bf 107}, 153 (1981).
\bibitem{e687-kkpi} E687 Collaboration, P.L. Frabetti {\em et al.}, Phys. Lett. B 
{\bf 351}, 591 (1995).
\bibitem{e691-3pi} E691 Collaboration, J.C. Anjos {\em et al.}, Phys. Rev. Lett. {\bf 62}, 
125 (1989).  
\bibitem{e687-3pi} E687 Collaboration, P.L. Frabetti {\em et al.}, Phys. Lett. B 
{\bf 407}, 79 (1997).
\bibitem{penni2} M.R.~Pennington, in {\it Proceedings of the Workshop on Hadron Spectroscopy},
Frascati Physics Series Vol. XV (Laboratory Nazionali de Frascati, Frascati (Roma), Italy, 1999),
p. 95.
\bibitem{torn3} N.~T\"ornqvist, in {\it Proceedings of the Workshop on Hadron Spectroscopy},
Frascati Physics Series Vol. XV (Laboratory Nazionali de Frascati, Frascati (Roma), Italy, 1999),
p. 237, and N.~T\"ornqvist,  hep-ph/0008135.
\bibitem{sigma-exp} WA102 Collaboration, D. Barberis {\em et al.}, Phys. Lett. B 
{\bf 453}, 316 (1999); CLEO Collaboration, D.M.~Asner {\em et al.}, Phys. Rev. D {\bf 61}, 
012002 (2000); GAMS Collaboration, D. Alde {\em et al.}, Phys. Lett. B {\bf 397 }, 350 (1997).
\end{thebibliography}
\end{document}